\documentstyle[prd,preprint,aps,epsfig,floats]{revtex}
\begin{document}

\tightenlines
\input epsf.tex
\def\DESepsf(#1 width #2){\epsfxsize=#2 \epsfbox{#1}}
\draft
\thispagestyle{empty}
\preprint{\vbox{\hbox{OSU-HEP-01-04} \hbox{UMDPP-01-62} \hbox{CTP-TAMU-20-01}
\hbox{June 2001}}}
\title{\Large \bf Solving the Strong CP and the SUSY Phase
Problems with Parity Symmetry}

\author{\large\bf K.S. Babu$^{(a)}$, B. Dutta$^{(b)}$ and R.N. Mohapatra$^{(c)}$}

\address{(a) Department of Physics, Oklahoma State University\\
Stillwater, OK 74078, USA \\}

\address{(b) Center for Theoretical Physics, 
Department of Physics\\ Texas A \& M University,
College Station, TX 77843-4242, USA \\}

\address{(c) Department of Physics, University of Maryland\\
College Park, MD 20742, USA}

\maketitle

\thispagestyle{empty}

\begin{abstract}

We propose a simultaneous solution to the strong CP problem and the SUSY phase 
problem based on parity symmetry realized when the supersymmetric standard 
model is embedded into a left--right symmetric framework at a scale near
$2 \times 10^{16}$ GeV, as suggested by neutrino masses and gauge coupling
unification.  In this class of models, owing to parity,
SUSY contributions to the muon anomalous magnetic moment can be naturally large
without conflicting with the EDM of the electron and the neutron.  The strong 
CP violation parameter $\bar{\theta}$ is zero at the tree level, also due to parity (P), 
but is induced due to P--violating effects below the unification scale.  
We estimate the induced $\bar{\theta}$ to be $\leq 10^{-16}$, if we adopt a 
constrained supersymmetric spectrum with universal scalar masses.
In the more general SUSY breaking scenario, after imposing flavor 
changing constraints, we find $\bar{\theta} \sim (10^{-8}-10^{-10})$, which is 
compatible with, but not much below the present limit on neutron EDM.  We also 
argue that potential non--perturbative corrections to $\bar{\theta}$ from
quantum gravitational effects are not excessive in these models. 

\end{abstract}

\newpage

\section{Introduction}

One of the major problems of the Standard Model is a lack of understanding of
the CP violating parameter $\bar{\theta}$ characterizing the QCD sector of
the Lagrangian \cite{kim}. This parameter originates from the periodic vacuum 
structure of QCD and leads to an electric dipole moment (EDM) of the neutron,
$d_n \sim (5 \times 10^{-16})~ \bar{\theta}$ e-cm.  The current 
experimental limit, $d_n \leq 6.3 \times 10^{-26}$ e-cm \cite{dn}, then
implies that $\bar{\theta} \leq 10^{-10}$.  Why a fundamental parameter
of the theory,
$\bar{\theta}$, is so small compared to its natural value of order unity is the 
strong CP problem.  A resolution of this problem is expected to provide
important clues to the nature of new physics beyond the Standard Model.

When the Standard Model is embedded into its minimal supersymmetric
extension (MSSM) in order to solve the quadratic divergence problem associated 
with the Higgs boson mass, one runs into another CP problem -- the SUSY phase 
problem,
or the SUSY CP problem.  This problem owes its origin to the 
complex phases associated with the parameters in the soft SUSY--breaking sector
of the theory -- the $\mu$ term, gaugino masses, trilinear $A$ terms, etc.
Exchange of supersymmetric particles in loops will induce electric
dipole moments (EDMs) for the neutron
($d_n$) and the electron ($d_e$) proportional to these SUSY phases.  $d_n$
induced through gluino and squark exchange can be estimated to be $d_n \approx
(1\times 10^{-23})$ e-cm $(300~{\rm GeV}/{M_{\rm SUSY}})^2
\sin\phi_{\rm SUSY}$, where $\phi_{\rm SUSY}$ is a typical SUSY CP phase
parameter
and $M_{\rm SUSY}$ is the gluino/squark mass.  (For this estimate
we used $\tan\beta = 5$ and set the $\mu$ parameter and the gluino mass equal
to the squark mass.)  In order to be compatible with the experimental limit
on $d_n$, the SUSY phase must obey
$\phi_{\rm SUSY} \leq (5 \times 10^{-3}-5 \times 10^{-2})$ for
$M_{\rm SUSY} = 300~{\rm GeV}-1 ~ {\rm TeV}$, unless the gluino contribution
is precisely canceled
by some other diagrams \cite{cancel}.  Why $\phi_{\rm SUSY}$ is so
small, while the corresponding CP violating phase in the CKM matrix is of order 
one (in order to explain the observed CP violation in the K meson system), is 
the SUSY phase problem.

The SUSY phase problem becomes more acute if we attribute the 
recently reported  2.6 sigma discrepancy in the muon anomalous
magnetic moment ($a_\mu$) measurement \cite{bnl}
to supersymmetry.  Exchange of charginos/neutralinos and 
sleptons can account for the observed discrepancy in $a_\mu$, provided
that the masses
of these particles are not more than about $200-400$ GeV.  Now, 
if we replace the external muons in the diagram responsible for $a_\mu$
by electrons, a large EDM for the electron will result, 
if the relevant SUSY phases are of order one.  For example, if the SUSY
contribution to $a_\mu$  is $40 \times 10^{-10}$, $d_e$ can be estimated
by scaling of the lepton mass to be $d_e \simeq (1.8 \times
10^{-24})\sin\phi_{\rm SUSY}$
e-cm.  The current experimental limit on $d_e$, viz., $d_e \leq 4.3
\times 10^{-27}$ e-cm \cite{de} would require 
$\phi_{\rm SUSY} \leq 2 \times 10^{-3}$.  This bound will be tightened even
further with the expected improvement in the limit on $d_e$ by about a
factor of 2 \cite{demille}.

Solving the strong CP problem as well as the SUSY phase problem are therefore  
major challenges facing the (supersymmetric) Standard Model.  

Simple solutions to the strong CP problem can be found by postulating new
symmetries of Nature, the most popular one being the Peccei--Quinn $U(1)_A$ 
symmetry \cite{pq}.  Consistent implementation of this symmetry requires
the existence
of a new ultralight particle, the axion, which has eluded experimental
searches so far.  Combined laboratory, astrophysical and cosmological limits
constrain the scale of Peccei--Quinn symmetry breaking to be in a narrow
window, $f_a \sim (10^{11} - 10^{12})$ GeV \cite{kim}.  On the theoretical 
side, global symmetries such as $U(1)_A$ have come under suspicion since 
it is believed
that nonperturbative quantum gravitational effects will violate all global
symmetries.  If so, quantum gravity would destabilize the axion solution 
\cite{holman} unless one allows for extreme fine--tuning of parameters -- the
very problem one set out to avoid in postulating the new
symmetry.\footnote{For example,
quantum gravity can induce a dimension 5 operator in the scalar potential
$g |\Phi|^4(\Phi + \Phi^*)/M_{\rm Pl}$, where $g$ is a dimensionless coupling,
and $\Phi$ is the scalar field responsible for $U(1)_A$ symmetry breaking.  Such
a term will induce a nonzero $\bar{\theta}$ given by $\bar{\theta} \sim
g f_a^5/(\Lambda_{QCD}^4 M_{\rm Pl}) \simeq 10^{40}g$.  The resulting constraint
on the coupling $g$ from neutron EDM is quite severe: $g \leq 10^{-50}$.}  
In any case, this solution has nothing to offer to the second CP problem,
the SUSY phase problem.
 
An alternative to the axion solution to the strong CP problem is
parity invariance \cite{goran} realized at a momentum scale $v_R$ much above 
the weak scale.\footnote{For models with CP invariance, see
\cite{nb}.  Related applications in SUSY context has been discussed in
Ref. \cite{dine}.}  If the Standard Model is embedded into a left--right
symmetric gauge structure at $v_R$, parity (P) invariance can
be consistently imposed on the Lagrangian.  In this case, the $\theta$
term in the QCD Lagrangian, $g^2/(32 \pi^2) \theta G \tilde{G}$, will be
zero since it violates parity.  The physical parameter $\bar{\theta}$
involves also the phase of the fermionic determinant, and is given by
\begin{equation}
\bar{\theta} = \theta +{\rm Arg}\{{\rm Det}(M_u M_d)\} - 3 {\rm Arg}(M_{\tilde{g}})~.
\end{equation}
Here $M_{u,d}$ are the up--quark and the down--quark mass matrices and
$M_{\tilde{g}}$ is the gluino mass.  Owing to parity invariance, the matrices
$M_u$ and $M_d$ become hermitian, and the gluino mass becomes real.  As
a result, $\bar{\theta} = 0$ at tree level in this class of models.  
These models would thus have the potential to solve the strong CP
problem.  The fact that low energy weak
interactions do not respect parity symmetry means that one must do additional
work to see if $\bar{\theta}$ induced through quantum effects
is sufficiently small. As we shall explicitly demonstrate in this paper,
this is often the case.  

The purpose of this note is to provide a simultaneous solution to the
strong CP problem and the SUSY phase problem
using parity symmetry.  We shall demonstrate by
explicit calculation that the induced $\bar{\theta}$ in these models is 
well within the experimental limit, if a constrained supersymmetric spectrum 
is adopted  with universal squark masses and proportional $A$ terms.
Even in the more general scenario for supersymmetry breaking, we shall see 
that $\bar{\theta}$ is in the acceptable range of $\sim 10^{-10}-
10^{-8}$, after imposing flavor changing constraints.  Parity invariance
also makes the phases of the SUSY breaking parameters naturally small.
These phase parameters are zero at the scale $v_R$ and their induced values
at the weak scale through quantum corrections are well within the experimental
limits arising from the neutron and the electron EDM.  Thus this class of models
can naturally explain the observed discrepancy in the muon anomalous
magnetic moment
$a_\mu$, without inducing unacceptably large EDM for the neutron and the
electron.  

In the class of models presented here the scale of parity restoration
is in the range $v_R \sim 10^{14}-10^{16}$ GeV.  This is close to the
grand unification
scale where the three gauge couplings of the Standard Model are observed to
unify in a supersymmetric context.  The left--right symmetric gauge structure 
and the numerical value of the
scale $v_R$ are independently suggested by experimental evidence for 
neutrino masses:
Small neutrino masses arising through the seesaw mechanism \cite{seesaw}
is natural in
the left--right framework, and the scale $v_R$ is consistent with the inferred 
value of $\nu_\tau$ mass from atmospheric neutrino oscillations.

Parity symmetry as a possible solution to the strong CP problem 
in supersymmetric contexts has been studied
in earlier papers \cite{ravi}.  The models presented here
are significantly improved versions in this regard.  In earlier works
\cite{ravi},
it was found that a consistent solution to the strong CP problem required
the scale of parity restoration to be in the multi--TeV range, which
would appear to not go well with neutrino masses and gauge coupling 
unification, unlike in the models presented here.
As we shall see, potential non--perturbative
corrections to $\bar{\theta}$ from quantum gravity are under control
here, even with $v_R$ near the unification scale.  
Parity as a solution to the SUSY phase problem has also
been studied in earlier papers \cite{rasin,bdm1}, where it has been shown
that the EDMs of the neutron and the electron remain very
small in the parameter space that fits  $\epsilon$ and $\epsilon'$ in the
Kaon system. Unlike in these earlier works which
preferred a non--KM mechanism for the Kaon CP violation, the models here
allow for the conventional KM CP violation.  This is 
facilitated by a novel realization of the doublet--doublet splitting --
the mechanism that makes one pair
of Higgs(ino) doublets light and all other pairs heavy, 
so that at low scale the spectrum
of the theory is identical to that of the MSSM.  Major differences of our
models compared to the MSSM are that here  (i) SUSY phases are naturally
small, (ii) the strong CP problem is absent, and (iii) small neutrino
masses are naturally present.  

\section{Basic outline of the model} 

The basic framework of our model involves the embedding of
the MSSM into a minimal SUSY left--right gauge structure 
at a scale $v_R$ close to the GUT scale.  The electroweak
gauge group of the model is
$SU(2)_L\times SU(2)_R\times U(1)_{B-L}$ with the standard assignment of
quarks and leptons --
left--handed quarks and leptons ($Q,L$) transform as doublets of
$SU(2)_L$, while the conjugate right--handed
ones ($Q^c,L^c$) are doublets of $SU(2)_R$.  
The quarks $Q$ transform under
the gauge group as $(2,1,1/3)$ and $Q^c$ as $(1,2,-1/3)$, while the
lepton fields $L$ and $L^c$ transform as $(2,1,-1)$ and $(1,2,+1)$
respectively.  
The Dirac masses of fermions arise through their Yukawa
couplings to one or more Higgs bidoublet $\Phi(2,2,0)$.  
The $SU(2)_R\times U(1)_{B-L}$ symmetry is broken down to $U(1)_Y$ in the
supersymmetric limit 
by $B-L=\pm 1$ doublet scalar fields, the right--handed doublet denoted by 
$\chi^c(1,2,-1)$ accompanied by its left--handed partner 
$\chi(2,1,1)$.  Anomaly cancellation requires the presence of
their charge conjugate fields as well, 
denoted as $\bar{\chi^c}(1,2,1)$ and
$\bar{\chi}(2,1,-1)$. The vacuum expectation values (VEVs) 
$\left\langle \chi^c \right\rangle  =
\left\langle \bar{\chi^c} \right \rangle = v_R$ break the left--right
symmetry group down to the  MSSM gauge symmetry. 

This embedding of MSSM into a left--right framework provides a simple
solution to the SUSY phase problem.  To see this, let us note the 
transformation of various fields under parity symmetry: 
$Q\leftrightarrow Q^{c*}, ~L \leftrightarrow
L^{c*}$, $\Phi \leftrightarrow \Phi^\dagger,~ \chi \leftrightarrow \chi^{c*},~
\bar{\chi} \leftrightarrow \bar{\chi^{c*}}$,~ $G \leftrightarrow G^*, ~B
\leftrightarrow
B^*$, $W_L \leftrightarrow W_R^*$ and $\theta \leftrightarrow \bar{\theta}$.  
Here $(G,~B,~W_{L,R})$ are the vector superfields associated with
$SU(3)_C$, $B-L$
and $SU(2)_{L,R}$ respectively, $\theta$ is the fermionic variable,
and the transformation applies to the matter
superfields as a whole.  Invariance under P makes the Dirac Yukawa couplings
of quarks and leptons and the associated SUSY breaking trilinear $A$--terms
hermitian. The gluino and the $B-L$ gaugino masses become real, the mass of the
bidoublet field $\Phi$ as well as the corresponding bilinear $B\mu$ term
are real, and $M_{L} = M_{R}^*$,
$M_{L,R}$ being the masses of the $SU(2)_{L,R}$ gauginos.  This resolves
the SUSY phase problem, since all the relevant SUSY
phases are zero at the scale of parity restoration \cite{rasin}. Renormalization
group extrapolation induces very small phases in the SUSY breaking parameters
of the MSSM, but these induced values are well consistent with experimental
limits arising from $d_n$ and $d_e$ \cite{bdm1}.  It is worth noting that
this solution to the SUSY phase problem will be valid
even in a general context of SUSY breaking, for example, without assuming
universality of scalar masses and proportionality between
the $A$ terms and the Yukawa couplings.  
Potential contributions to the EDM of the neutron and the electron will be
proportional to the diagonal entries
of the respective $A$ matrix or the squark/slepton mass--squared matrix.  
Both matrices being hermitian, these contributions vanish
above the scale $v_R$.  Since the gaugino masses are all real (assuming
gaugino mass unification that occurs in various scenarios of SUSY breaking
even without a unifying group, or if the left--right
gauge theory is embedded into a higher symmetry group such as $[SU(3)]^3$,
$SO(10)$ or $E_6$), $d_n$ and $d_e$ proportional to the phases of the
gaugino masses also vanish above $v_R$.  

\subsection{The doublet--doublet splitting mechanism}

In order to make the supersymmetric left--right gauge theory
fully realistic, a mechanism should be found that keeps one pair of
Higgs doublets light at the weak scale (to be identified with the $H_u$ and 
$H_d$ fields of MSSM) and any remaining pairs of
Higgs doublets superheavy at the scale 
$v_R$.  The simplest possibility would appear to be to
introduce just a single Higgs 
bidoublet $\Phi(2,2,0)$
which gets a mass of order SUSY breaking scale.  However, 
this is not the minimal scenario from the effective low energy
point of view, since in that case the up and down quark mass matrices
become proportional, leading to vanishing quark mixings at tree 
level.\footnote{Realistic quark
mixings can be induced through the gluino and chargino loops, provided that 
the trilinear $A$ terms have a flavor structure different from that of the
Yukawa couplings \cite{bdm1,bdm2}.  
Consistency with flavor changing processes would require
that $\tan\beta$ be not too large \cite{bdm1}, $\tan\beta \leq 10$,
which excludes the simplest scenario where $\tan\beta = m_t/m_b \simeq 60$.  
Values of $\tan\beta$ smaller than $m_t/m_b$ may be obtained
if the $\Phi$ field mixes with some other superheavy doublets
of the theory in a parity violating manner (if such mixings conserve P,
$\tan\beta = m_t/m_b$ will prevail).  In this case
the effective $\mu$ and $B\mu$ terms are
potentially complex, which would spoil the
solution to the SUSY phase problem based on parity.
To maintain this solution, in earlier works \cite{bdm2}
we assumed invariance under charge conjugation symmetry C, in addition
to P, which allows for reality of the effective $\mu$ and $B\mu$ terms.
In such a scenario the CKM phase will be zero due to C invariance and
the observed CP violation in the Kaon system is explained
through supersymmetric gluino/squark diagrams.}

In this subsection we shall present a new mechanism for doublet--doublet
splitting.  It involves two bidoublet Higgs fields and is achieved without
any  fine--tuning of parameters.  The solution to the SUSY phase
problem is preserved based on parity symmetry alone.  As we shall elaborate
further in the next subsection, in this scenario
the quark mixings arise naturally at tree level, $\tan\beta$ can be 
smaller than $m_t/m_b$,
symmetry breaking
occurs at the renormalizable level without any pseudo--Goldstone
bosons, and neutrino masses are correctly reproduced with $v_R$ near
the unification scale.  Furthermore, the effective low energy theory is 
just the MSSM, with a natural understanding of the weak scale value of
the $\mu$ parameter which remains real.  

Consider the following form of the superpotential 
involving two bidoublet field $\Phi_a$, $a=1,2$ and the left--handed
($\chi+\bar{\chi})$ and  and the
right--handed ($\chi^c+\bar{\chi^c}$) doublets of the theory:
\begin{equation}
W~=~ \lambda_a \chi \Phi_a \chi^c +\lambda'_a \bar{\chi} \Phi_a\bar{ 
\chi^c} + M\chi\bar{\chi} + M^*\chi^c\bar{\chi}^c ~.
\end{equation}
Parity invariance makes the couplings $\lambda_a$ and $\lambda_a'$ real,
since $\Phi \rightarrow \Phi^\dagger$ and $\chi \rightarrow \chi^{c*}$
under P.  The mass term $M$ and the VEV $\left\langle \chi^c\right\rangle
= \left\langle \bar{\chi^c}\right\rangle = v_R$ are complex in general.
After $SU(2)_R$ breaking, this superpotential leads to 
a mass matrix for the doublets given in Eq. (3). We use a notation in which the 
rows denote $(\phi_{u1}, \phi_{u2}, \chi)$ fields and the columns denote
$(\phi_{d1}, \phi_{d2}, \bar{\chi})$ fields where $\phi_{u1}$ and 
$\phi_{d1}$ are the up and down type Higgs doublets from $\Phi_1$, etc.
\begin{eqnarray}
M_{DD} = \left( \begin{array}{ccc}
0 & 0 & \lambda_1 v_R \\ 0 & 0 & \lambda_2 v_R \\
\lambda_1' v_R & \lambda_2' v_R & M 
\end{array} \right)~.
\end{eqnarray}

This  mass matrix leaves one pair of Higgs doublets massless, while
giving mass of order $v_R$ to the second pair. 
Since $\lambda_a$ and
$\lambda_a'$ are real, the effective $\mu$ term of the light doublets becomes
real.  To see this, observe that
the low energy MSSM doublets
are given by $H_u = \cos\alpha_u ~\phi_{u1} + \sin\alpha_u~ \phi_{u2}$ and $H_d
= \cos\alpha_d ~\phi_{d1} + \sin\alpha_d~\phi_{d2}$, where $\tan\alpha_u =
\lambda_1/\lambda_2$ and $\tan\alpha_d = \lambda_1'/\lambda_2'$.  Note that
$H_u$ and $H_d$ are $real$ linear combinations of $\Phi_a$, which helps in
inducing a real $\mu$.  
The superpotential of Eq. (2) by itself
does not lead to a $\mu$ term, which gets
induced only after SUSY breaking.  There are two sources that induce the $\mu$
term:  
\begin{enumerate}
\item Kahler potential terms
of the form $\lambda_{ab}''\int d^4\theta 
\frac{Z^{*}}{M_{\rm Pl}}{\rm Tr}(\Phi_a\Phi_b)$,
where
$Z$ is a gauge singlet whose $F_Z\neq 0$ breaks
supersymmetry. We 
assume that $Z$ is parity even, i.e., $Z\rightarrow Z^*$ under P.  The coupling
matrix $\lambda''$ is therefore hermitian.  
After supersymmetry breaking this term will lead to a real $\mu$-term, as desired.
This also provides a reason why the $\mu$-term is of order of the
electroweak scale\cite{giudice}.
The $B\mu$ term arises from the term $\int d^4\theta
\frac{ZZ^*}{M^2_{\rm Pl}}~{\rm Tr}(\Phi_a\Phi_b)$, which 
is also real due to parity.  

\item A second mechanism that generates real $\mu$ and
$B\mu$ terms makes use of the superpotential couplings involving a
visible sector singlet $S$: $W \supset  \kappa S (e^{i \xi}
\chi^c \bar{\chi^c} + e^{-i\xi} \chi \bar{\chi} - M^2)$.
Such a coupling can break the left--right gauge symmetry down to the MSSM
symmetry
at the renormalizable level without leaving any pseudo--Goldstone bosons.
Owing to parity, under which $S \rightarrow S^*$, 
the parameters $\kappa, M^2$ are real in this superpotential
coupling.  The field $S$ also has the following coupling to the bidoublets
$\Phi_a$: $W \supset \mu_{ab} {\rm Tr}(\Phi_a\Phi_b)S$.  
In the SUSY
limit, $\left\langle S \right\rangle = 0$, $\left\langle \chi^c \right\rangle =
\left\langle \bar{\chi^c}\right\rangle = v_R$, $\left\langle \chi \right\rangle=
\left\langle \bar{\chi} \right\rangle = 0$, which breaks parity
spontaneously.  $S$ pairs up with the neutral 
component $(\chi^c + \bar{\chi^c})/\sqrt{2}$ to form a multiplet that has mass 
$\sqrt{2}\kappa v_R$.  After SUSY breaking, the coupling $\int d^2\theta {\chi^c
\bar{\chi^c} S Z \over M_{\rm Pl}}$ will induce a tadpole in 
Re($S$) scalar of order $v_R^2 m_{3/2}$.
A VEV $\left\langle {\rm Re}(S) \right\rangle \sim m_{3/2}$ will result,
that provides
a real $\mu$ term for the bidoublet fields.  
It is crucial to note that by redefining the $\chi^c$ filed, the coefficient of
the tadpole in $S$ can be made real without introducing any phases
elsewhere (see
Eq. (4) below).  If the imaginary component of $S$ had also a tadpole,
$\left\langle {\rm Im}(S)\right\rangle \neq 0$,
which will lead to an effective complex $\mu$ term.  
\end{enumerate}

\subsection {The full Lagrangian}

Now let us implement the doublet--doublet splitting mechanism just
described.  We shall see that there is a discrete anomaly--free
$Z_4$ R--symmetry that achieves this goal within a minimal version
of the left--right model (viz., using two bidoublets $\Phi_a$,
one left--handed ($\chi$) and one right--handed ($\chi^c$) $SU(2)$
doublets along with
their conjugate ($\bar{\chi}+\bar{\chi^c}$) and the singlet $S$).  
Their transformations under P has been given earlier, with $S \rightarrow
S^*$.  All the desired terms,
including the Majorana mass terms for the right--handed neutrinos are allowed
by this $Z_4$, and the unwanted terms that can potentially make
the magnitude of $\mu$ too large, or induce excessive CP phases
to upset the strong CP and the 
SUSY phase solutions will be prevented.  
The $Z_4$ is broken at the scale $v_R$, but a $Z_2$
remnant remains, which is identified as the usual R parity
of the MSSM.  This $Z_2$ will guarantee the stability of the proton.

Under the $Z_4$ R symmetry, the superpotential changes sign ($ W \rightarrow
-W$), as do $d^2\theta$ and $d^2\bar{\theta}$.  
The gaugino fields transform as $\lambda_a \rightarrow -\lambda_a$,
quarks and leptons
are even, $\Phi_a: -1,~\chi: i,~\bar{\chi}: -i$, $\chi^c: -i,~ \bar{\chi^c}: +i,
~S: -1$.  

The gauge invariant superpotential consistent with this $Z_4$ R symmetry is

\begin{eqnarray}
W &=& h_aQ \Phi_aQ^c + h_a'L \Phi_aL^c  + \lambda_a\chi \Phi_a \chi^c +
    \lambda_a'\bar{\chi} \Phi_a \bar{\chi^c} + 
(f L L \chi \chi +f^*L^c L^c \chi^c \chi^c)/M_{\rm Pl} +  \nonumber \\
&~& \kappa S (e^{i \xi}\chi^c \bar{\chi^c}+
     e^{-i\xi}\chi \bar{\chi} + aS^2 - M^2)  + \mu_{ab}{\rm
Tr}(\Phi_a\Phi_b) S ~.
\end{eqnarray}

This superpotential induces tree level CKM mixings since the light
MSSM doublets $H_{u,d}$ are parity asymmetric linear combinations
of the two bidoublets.
The $f$ couplings give rise to Majorana masses for $\nu_R$
of order $v_R^2/M_{\rm Pl}$.  For $v_R \sim 10^{14}-10^{16}$ GeV, the
magnitude of the light neutrino masses are in the right range to explain
the atmospheric and the solar neutrino oscillation data.  
$f$ could have its origin in quantum gravity, but it could also
arise from integrating out singlets which have masses of order $M_{\rm
Pl}$, e.g.,
through $(L N \chi + L^c N^c \chi^c)$ couplings where
$(N, N^c$) are the singlets with $Z_4$ charges $(i,-i)$. Their Majorana masses
$[N^2 + N^{c^2}]$ preserve the $Z_4$ symmetry.

In the SUSY limit, we have $\left\langle S\right\rangle = 0,
 \left\langle \chi^c\right\rangle = \left\langle
\bar{\chi^c}\right\rangle = M$
with all other fields having zero VEVs. As noted earlier, 
after SUSY breaking, the real component of $S$ gets an induced VEV 
of order $m_{3/2}$.  (Note that the phase in the $\chi^c \bar{\chi^c}S$ scalar
coupling can be made real by redefining $\chi^c$ field.  This redefinition
does not induce new phases anywhere else.  The tadpole in $S$ that is
induced after SUSY breaking is therefore real, making only Re($S$) to be
nonzero.) That gives a hermitian $\mu_{ab}$ terms through the
last couplings of Eq. (4), or to real $\mu$ parameter.  
We can also have
the coupling $\lambda_{ab}''\int d^4\theta {[{\rm Tr}(\Phi_a\Phi_b)] Z\over
M_{\rm Pl}}$ where $Z$ is the spurion field that breaks SUSY.  
This also leads to real $\mu$ term of the right order of magnitude.  (Note that
$Z$ is parity even, $Z \rightarrow Z^*$ under P.  $F_Z$ is then expected to
be real, which would leave parity unbroken.  For example, in the Polonyi
model of
hidden sector SUSY breaking, $W \supset \mu^2(Z + \beta)$, where $\mu^2$ is
real due to parity.  $F_Z = \mu^2$ is therefore real.  We anticipate the reality
to $F_Z$ to hold even in a more general scenario for SUSY breaking.)

The superpotential of Eq. (4)  
reproduces the doublet--doublet mixing matrix of Eq. (3).
In Eq. (3), the (3,3) entry is of order $m_{3/2}$ now,
being proportional to $\left\langle S \right\rangle$.  
It does not
correspond to any new particle having mass of order $m_{3/2}$, since
$\chi$ pairs with
the heavy doublet in $\Phi_a$ and has a mass of order $v_R$.

We shall now show that the
$Z_4$ R symmetry is an anomaly free discrete gauge symmetry \cite{ir}.  This
makes it aesthetically more pleasing, as it may have its origin in
a true gauge symmetry.  It also protects the Lagrangian from receiving
uncontrollable quantum gravitational correction.
To see the anomaly freedom, let us assume that the $Z_4$ arose from a 
true gauge $U(1)_R$ symmetry.  The $U(1)_R$ then should be anomaly free.
If the $U(1)_R$ symmetry is broken at the Planck scale by scalar VEVs that are
integer multiple of $\pm 4$ (upto an overall normalization factor),
a residual $Z_4$ symmetry will survive.  The $U(1)_R$ anomaly cancellation
will in general require introduction of additional fermionic fields.  
The crucial question is then 
if these extra fields can be removed from the low energy spectrum by
giving them $Z_4$ invariant masses.  To address this, 
let us embed the $Z_4$ into
a $U(1)_R$ in the obvious fashion, by assigning $U(1)_R$ charges as follows:
(We display the R charge of the superfield, which is the same for
the scalar component,
but the fermionic component will have its R charge shifted by $-1$.)
Quarks and leptons: 0, $\chi: +1, ~\chi^c: -1, ~\bar{\chi}: -1,~
\bar{\chi^c}: +1,~ \Phi_a: +2, ~S: +2$.  The superpotential $W$
has R charge of $+2$, and the gauginos have R charge of $+1$.

With this assignment, one can compute all the mixed anomaly coefficients:
\begin{eqnarray}
SU(3)_C^2 \times U(1)_R :    3 - 2 N_g \nonumber\\
SU(2)_L^2 \times U(1)_R:   3 - 2 N_g \nonumber \\
SU(2)_R^2 \times U(1)_R:   3 - 2 N_g \nonumber \\
U(1)_{B-L}^2 \times U(1)_R: -3 - 2 N_g 
\end{eqnarray}

\noindent Here $N_g = 3$ is the number of
generations.  The 3 in first term arises from gluino loop.  The
3 in second term is from Wino (+2) and Higgsino (+2 from
the two bidoublets and $-1$ from $\bar{\chi}$).  We have used the
conventional $SO(10)$~ $B-L$ normalization, $\sqrt{3/2} (B-L)/2$
being the normalized generator.  

Since all the nonabelian mixed anomaly coefficients are equal, 
we can try to cancel them
by the Green-Schwarz mechanism \cite{gs}.  We shall make the abelian
mixed anomaly coefficient to be also equal to
facilitate this.  This can be achieved
by adding a pair of singlets
which are $(2,3) + (-2,3)$ under $(B-L, R)$.  Then the
$U(1)_{B-L}^2 \times U(1)_R$
anomaly also becomes $(3 - 2 N_g)$.  A mass term for these singlet
fields will have R charge of $+6$ (scalar component), so it must
be accompanied by a Higgs field with charge $-4$.  That
breaks $U(1)_R$ to $Z_4$, as desired.  

Finally, there is $U(1)_R^2 \times U(1)_{B-L}$ anomaly, which has
a coefficient $-8 \sqrt{3/2}$
in this model.  
We use the overall R--normalization (level of $U(1)_R$) to make
this equal to the other anomaly coefficients.  This normalization factor is
then found to be $(1/4)\sqrt{3/2}$ -- very similar to $B-L$
normalization.
The $[U(1)_R]^3$ anomaly also has the same coefficient (equal to $-3$
for $N_g = 3$) if some singlet fields contribute $+24$ in the cubic
anomaly.  3 singlets with fermionic R-charge of $+2$ will
do this job.  Mass terms for these singlets will carry
R-charge of $+6$ (scalar component), so to make it $+2$,
we must multiply by a Higgs with R--charge of $-4$.  Again,
we see that the $Z_4$ is left unbroken by this Higgs.
This way of canceling anomalies is not unique, but to
establish that $Z_4$ is discrete anomaly free, any one example
would suffice.

\section{Calculation of induced $\bar{\theta}$}

As noted earlier, parity invariance implies that the QCD Lagrangian parameter 
$\theta =0$, the gluino mass is real and the quark mass matrices
$M_{u,d}$ are hermitian at tree level.  Therefore $\bar{\theta}=0$ at
tree level.  Since parity is broken at $v_R$, a nonzero value of
$\bar{\theta}$ will be induced at the weak scale through renormalization
group extrapolation below $v_R$.  We shall estimate this
induced $\bar{\theta}$ in two scenarios for SUSY breaking.  The first is
the constrained MSSM scenario where the squark masses are degenerate at
the unification scale with the trilinear $A$ matrices and the corresponding
Yukawa coupling matrices being proportional.
The second scenario has
a more general SUSY breaking spectrum without universality or proportionality,
but the experimental constraints
arising from flavor changing processes will be imposed.  We first turn to
the correction to $\bar{\theta}$ arising from the non--hermiticity of the
Yukawa coupling matrices which applies to both these scenarios.

\vspace*{0.1in}
\noindent {\bf 1. $\delta\bar{\theta}$ from non--hermiticity of the
Yukawa coupling matrices:}
\vspace*{0.1in}

At the scale $v_R$, the up and down Yukawa coupling matrices are hermitian
owing to parity.  The VEVs of the MSSM fields $H_{u,d}$ are real since
the $B\mu$ term is real, so that the quark mass matrices $M_u$ and $M_d$ 
are hermitian
at $v_R$.  They will develop non--hermitian components at the weak scale, 
owing to renormalization group evolution below $v_R$.  
The induced $\bar{\theta}$ will have the general structure given
by
\begin{eqnarray}
\delta \bar{\theta} = {\rm Im}{\rm Tr}[\Delta M_u M^{-1}_{u} + \Delta M_d
M^{-1}_{d}] -3 {\rm Im}(\Delta M_{\tilde{g}} M_{\tilde{g}}^{-1})
\end{eqnarray}
where  $M_{u,d,\tilde{g}}$ denote the tree level contribution to the
up--quark matrix,
down--quark matrix  and the gluino mass respectively, and $\Delta
M_{u,d,\tilde{g}}$
are the loop corrections.  To estimate the corrections from $\Delta M_u$
and $\Delta M_d$, we note that the
beta function for the evolution of $Y_u$ below $v_R$
is given by $\beta_{Y_u} = Y_u/(16 \pi^2)(3 Y_u^\dagger Y_u
+ Y_d^\dagger Y_d + G_u)$ with the corresponding one for $Y_d$ 
obtained by the interchange $Y_u
\leftrightarrow Y_d$ and $G_u \rightarrow G_d$.  Here $G_u$ is 
a family--independent contribution arising from gauge bosons and
the Tr$(Y_u^\dagger Y_u)$ term. The $3Y_u^\dagger Y_u$ term and the $G_u$ term
cannot induce non--hermiticity in $Y_u$, given that $Y_u$ is hermitian
at $v_R$.  The interplay of $Y_d$ with $Y_u$ will however induce deviations
from hermiticity.  Repeated iteration of the solution with $Y_u \propto
Y_uY_d^\dagger
Y_d$  and $Y_d \propto Y_dY_u^\dagger Y_u$ in these equations
will generate the following structure:
\begin{equation}
\delta \bar{\theta} \simeq 
\left({{\rm ln}(M_U/M_W) \over 16 \pi^2}\right)^4\left[
c_1 {\rm Im}{\rm Tr}\left(Y_u^2 Y_d^4 Y_u^4 Y_d^2 \right)
+c_2 {\rm Im}{\rm Tr}\left(Y_d^2 Y_u^4 Y_d^4 Y_u^2 \right)\right]~,
\end{equation}
where $M_U$ is the unification scale.  
Here $c_1$ and $c_2$  are order one coefficients which are not equal since
the flavor independent parts $G_u$ and $G_d$ are not the same for the evolution
of $Y_u$ and $Y_d$ (hypercharge gauge couplings and the tau lepton Yukawa
couplings differentiate the two.)  These contributions
to $\delta \bar{\theta}$ are very high order in the Yukawa couplings since
the trace of products of two hermitian matrices, having the form Im
Tr($Y_u^n Y_d^m
Y_u^p Y_d^q..$) contains an imaginary piece only at this order.
To estimate
the induced $\bar{\theta}$, we choose a basis where $Y_u$ is diagonal, $Y_u = D$
and $Y_d = V D' V^\dagger$ where
$D_u v_u = {\rm diag}(m_u,~m_c,~m_t)$, 
$D_dv_d = {\rm diag}(m_d,~m_s,~m_b)$ with $V$ being the CKM matrix.  The Trace 
of the first term in
Eq. (7) is then Im($D_i^2 D_k^4 D_j^{\prime^4} D_l^{\prime^2} V_{ij}V_{kl}V_{il}^*
V_{kj}^*)$.  The leading contribution in this sum is $(m_t^4 m_c^2 m_b^4 
m_s^2)/(v_{u}^{6}v_d^6){\rm Im}(V_{cb}V_{ts}V_{cs}^* V_{tb}^*)$.  
The second Trace in Eq.(7) is identical, except that it has an opposite sign.
Numerically then,
\begin{equation}
\delta \bar{\theta} \sim  3 \times 10^{-27} ({\tan\beta})^6 (c_1-c_2)
\end{equation}
where we have used the running quark masses at $m_t$ 
to be $(m_t, m_c, m_b, m_s) = (166, 0.6, 2.8, .063)$ GeV.  
Clearly, $\delta
\bar{\theta}$ is very small, even for $\tan\beta = 50$ its value is $10^{-16}$,
much below the experimental limit of $10^{-10}$ from neutron EDM.

Since the up(down)--quark mass matrix is a product of $Y_u(Y_d)$ and the VEV 
$v_u(v_d)$, the mass matrix can become complex if the VEV $v_u(v_d)$ is complex.
If the bilinear soft SUSY breaking parameter $B\mu$ becomes complex in the
process of evolution below $v_R$, this will happen.  By analyzing the RGE for
the $B\mu$ parameter, one sees that it involves traces of $(Y_u^\dagger
Y_u)$ and
$(Y_d^\dagger Y_d)$ or their products -- in the case of universal squark masses
and proportional $A$ terms
($A_u \propto Y_u, A_d \propto Y_d$).  We are again left with two hermitian
matrices $(Y_u, Y_d)$, with all other effective parameters being
real.  The imaginary
component of the trace that induces a phase in $B\mu$
is then given at lowest order by an expression analogous
to Eq. (7).  The estimate on $\delta \bar{\theta}$ is of the same order
as before, $\delta\bar{\theta} \sim 10^{-26} (\tan\beta)^6$.

\vspace*{0.1in}
\noindent{\bf 2. $\delta \bar{\theta}$ from finite correction to the quark
and the gluino masses:}
\vspace*{0.1in}

To compute $\delta \bar{\theta}$ arising from 
the finite corrections to the quark mass matrices and the gluino
mass (which are not contained in the RGE evolution), we must specify the
SUSY breaking spectrum.  
The simplest approximation is to assume
universality of scalar masses and proportionality of $A$-terms and
the respective Yukawa
couplings at the Planck scale. This can be justified in models such as the
ones with gauge mediated supersymmetry breaking\cite{martin}. In
this case, the whole 
theory at the weak scale is characterized by only two Yukawa coupling
matrices $Y_{u,d}$.  Furthermore, all other MSSM parameters are real in the
effective low energy theory below $v_R$.  
Because of this property it is very easy to estimate the
lowest order contribution to nonvanishing $\bar{\theta}$ in terms of the
coupling matrices. 

Consider the finite one loop corrections to the quark mass matrices.  A typical 
diagram involving the exchange of squarks and gluino is shown in Fig. 1.  
There are analogous chargino diagrams as well.  In Fig. 1, the crosses on the
$\tilde{Q}$ and $\tilde{Q^c}$ lines represent (LL) and (RR) mass insertions
that will be induced in the process of RGE evolution.  From this figure we
can estimate the
form for  $\Delta M_u = \frac{2\alpha_s}{3\pi} m^2_{\tilde{Q}} A_u
m^2_{\tilde{u^c}}$ where $\tilde{Q}$ is the squark doublet and
$\tilde{u^c}$ is the 
right--handed singlet up squark. Without RGE effects, the trace of this term
will be real, and will not contribute to $\bar{\theta}$.  
Looking at the RGE for 
$m^2_{\tilde{u^c}}$ upto two loop
order, we see that for the case of proportionality of $A_u$ and $Y_u$, 
$m_{\tilde{u^c}}^2$ gets corrections having the form
$m^2_0 Y^2_u$ or $m^2_0Y^4_u$ or $m^2_0
Y_uY^2_dY_u$. Therefore in $\Delta M_u M^{-1}_u$, the $M^{-1}_u$ always
cancels and we are left with a product of matrices of the form
$Y^n_uY^m_dY^p_uY^q_d\cdot\cdot\cdot$. A similar comment applies when we
look at the RGE corrections for $m^2_{\tilde{Q}}$ or $A_u$.  
If the product is hermitian, then
its trace is real. So to get a nonvanishing contribution to theta, we have
to find the lowest order product of $Y^2_u$ and $Y^2_d$ that is
non--hermitian\footnote{Similar
reasoning was used in the standard model and supersymmetric models in
earlier papers \cite{ellis}} and we get 
\begin{equation}
\delta \bar{\theta} =
\frac{2\alpha_s}{3\pi} \left({{\rm ln}(M_U/M_W) \over 16 \pi^2}\right)^4
\left(k_1{\rm Im} {\rm Tr}[Y^2_uY^4_dY^4_uY^2_d]+
k_2{\rm Im}{\rm Tr}[Y^2_dY^4_uY^4_dY^2_u]\right)
\end{equation}  
where $k_{1,2}$ are calculable constants.  The numerical estimate of this
contribution parallels that of the previous discussions, $\delta \bar{\theta}
\sim (k_1-k_2) \times 10^{-28} (\tan\beta)^6$.  The contributions from the
up--quark and down quark matrices tend to cancel, but since the $\tilde{d^c}$
and the $\tilde{u^c}$ squarks are not degenerate, $k_1 \neq k_2$ and the
cancellation is incomplete.

\begin{figure}[htb]
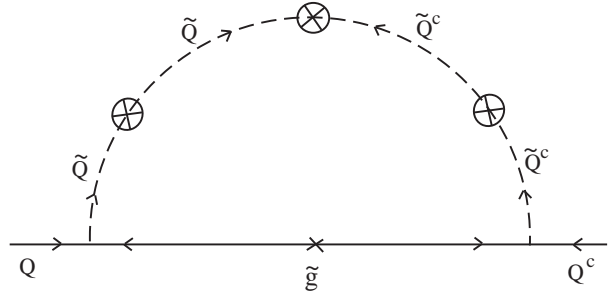

\centerline{ \DESepsf(strongcp.epsf width 8 cm) }
\bigskip
\bigskip
\caption {\label{fig1} One--loop gluino/squark exchange
diagram contribution to the quark mass matrix.  The crosses
on the scalar lines correspond to mass insertions.}
\end{figure}

In Fig. 2 we have displayed the one--loop contribution to the gluino mass
arising from the quark mass matrix.  Here again one encounters the imaginary
trace of two hermitian matrices $Y_u$ and $Y_d$, in the case of universality
and proportionality of SUSY breaking parameters.  Our estimate for $\delta
\bar{\theta}$ is similar to that of the quark mass matrix of Eq. (9).

\begin{figure}[htb]
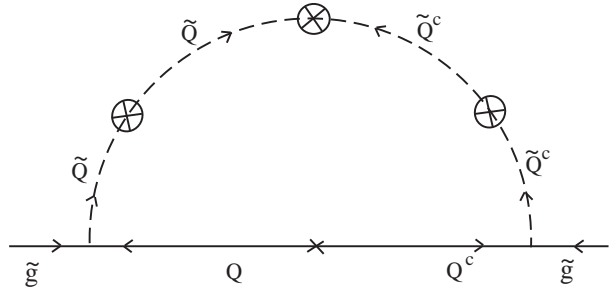

\centerline{ \DESepsf(strongcp1.epsf width 8 cm) }
\bigskip
\bigskip
\caption {\label{fig2} One loop diagram that inuduces a phase
in the gluino mass.}
\end{figure}

\vspace*{0.1in}
\noindent{\bf 3. Induced $\bar{\theta}$ with general SUSY breaking terms:}
\vspace*{0.1in}

In this subsection, we study the more  general SUSY breaking scenario where
soft SUSY breaking terms involving squarks is given by
\begin{equation}
{\cal L}_{SSB}~=~ \sum_{\phi=\tilde{Q}, \tilde{u^c},
\tilde{d^c}\cdot\cdot\cdot} \phi^{\dagger}m^2_{\phi}\phi
+\tilde{Q}A_uH_u\tilde{u^c} +\tilde{Q}A_dH_d\tilde{d^c} + h.c.
\end{equation}
For the model under study, at the $v_R$ scale, the constraint is
that
${m^2_{\tilde{Q}}}^T=m^2_{\tilde{u^c}}=m^2_{\tilde{d^c}}\equiv
~m^2_{\tilde{Q^c}}$ due to parity invariance. $A_{u,d}$
are arbitrary hermitian matrices, and the squark mass matrices 
can have non--trivial flavor structure.

In this case, the lowest order correction to $\delta \bar{\theta}$
from one loop contributions to quark masses (Fig. 1) is given by 
\begin{eqnarray}
 \delta \bar{\theta} \simeq \frac{2\alpha_S}{3\pi m^5_0} {\rm Im} {\rm Tr}
[m^2_{\tilde{Q}}A_fm^2_{\tilde{Q}}Y_f^{-1}]=0~
\end{eqnarray}
for $f=u,d$.  
This contribution vanishes since the matrix $m^2_{\tilde{Q}}A_f m^2_{\tilde{Q}}$ and
$Y_f^{-1}$ are both hermitian. The next leading contribution has the form
\begin{eqnarray}
 \delta \bar{\theta} \simeq \frac{2\alpha_Sv_{wk}}{3\pi
m^6_0}\frac{{\rm ln}(M_U/M_{SUSY})}{16\pi^2} {\rm Im}{\rm Tr}
[m^2_{\tilde{Q}}A_u m^2_{\tilde{Q}}Y_u]~.
\label{theta}
\end{eqnarray}
This contribution arises from Fig. 1 by inserting $m_{\tilde{Q}}^2Y_u^2$ 
arising from the RGE equations in one of the squark lines.  
Since this trace involves three arbitrary hermitian matrices, it is
not real in general.  To estimate this contribution, we have to make
some assumption about the non--universality in $m_{\tilde{Q}}^2$ and
the non--proportionality in $A$ and the Yukawa coupling matrix.  
As for the $A$ term, the most natural choice will be to assume that
it has the same hierarchical structure as the Yukawa couplings.  Such
a form would be suggested by flavor symmetries.  Thus, we shall take
$A_{23} \sim \epsilon A_{33}$, where $\epsilon$ is a small parameter,
of order $V_{cb} \sim1/30$.  Such a choice will guarantee that there
is no excessive FCNC processes mediated by squarks.  As for the squark
mass matrices $m^2_{\tilde{Q}}$, we take it to be approximately proportional
to a unit matrix, with correction terms that are not large.  This is as
suggested by nonabelian horizontal symmetries \cite{bmohap}.  The
leading contribution
from Eq. (12) to $\bar{\theta}$ arises when we use index $(3,2)$ for the
first $m_{\tilde{Q}}^2$, $(2,3)$ for $A_u$ and $(3,3)$ for the rest.  Using
$v_{wk}/m_0 \sim 1/5,~A_{33}/m_0 \sim 1/10,~A_{23} \sim 10^{-2}A_{33},~
(m_{\tilde{Q}}^2)_{23} \sim 10^{-2} m_0^2$, we find $\delta \bar{\theta} \sim
10^{-8}$.  This is a conservative estimate and yet it is encouraging that
we are close to the present upper limit on $\delta \bar{\theta}$ of $10^{-9}$ to
$10^{-10}$.  To be completely consistent with the neutron EDM limit, we
should have the relevant phase to be of order 0.1, or the off--diagonal
entries somewhat smaller than allowed by FCNC constraints.  Since such
departures from natural values need be only mild, we feel that this scenario
is also quite viable.  It is interesting that
in this scheme, $d_n$ is not much below the present experimental limit,
while $d_e$ is well below the current limit. 

Let us now address the contribution to $\bar{\theta}$ 
arising from the induced phase of the gluino
mass. The leading contribution (see Fig. 2) in this case is given by $\delta 
\bar{\theta}
\simeq \frac{2\alpha_S}{3\pi} {\rm Im}{\rm Tr} (A_uY_u)\frac{v_{wk}^2}{m_0^2
M_{\tilde{g}}}$. 
Without RGE running, this trace is real since $Y_u$ and $A_u$ are hermitian.
Allowing for RGE running, we estimate $\delta \bar{\theta} \simeq
\frac{2\alpha_S}{3\pi} \left[{\rm ln}(M_U/M_W)/(16 \pi^2)\right]
{\rm Im}{\rm Tr} (M_{\tilde{Q}}^2Y_uA_u)
\frac{v_{wk}^2}{m_0^4M_{\tilde{g}}}$.  Taking the (2,3) entries of
$M_{\tilde{Q}}^2$ and $A_u$ to be $10^{-2}$ times that of the respective
(3,3) entries, and with $A_{33} = m_0/10$, we arrive at $\delta \bar{\theta}
\simeq 10^{-8} - 10^{-10}$ for $v_{wk}/m_0 \sim 1/5$.  
This is again not far from the
present upper limit and with a mild fine--tuning of parameters,
of order 10\%,  one gets the desired solution to the strong CP 
problem.

\section{Planck scale corrections}

One interesting aspect of the model presented here is that it is quite
safe from potentially large corrections to $\bar{\theta}$ induced by quantum
gravity.  If it is assumed that the 
high scale parity conserving theory originates from a more
fundamental theory, one can expect nonrenormalizable operators 
in the theory suppressed by the mass scale associated with the 
fundamental theory. 
Such corrections to $\bar{\theta}$  will respect the gauge symmetry as well as
the anomaly free $Z_4$ discrete gauge symmetry.  We should ensure two 
things: (i) The effective $\mu$ term induced by quantum gravity is not
more than the weak scale, and (ii) The quantum gravity induced phases
which may not respect parity do not upset the solution to the strong
CP problem.  All other constraints, such as the solution of the SUSY
phase problem, will be automatically satisfied once these two are
taken care of.  

As for the magnitude and the phase of
the effective $\mu$ term, the most relevant higher dimensional
operator suppressed by Planck mass that is invariant under the
gauge symmetries and the $Z_4$ symmetry is $W \supset \kappa_{ab}{\rm Tr}(\Phi_a
\Phi_b)\chi^c \bar{\chi^c} S/M_{\rm Pl}^2$.  The magnitude 
of the resulting $\mu$ term is $\kappa_{ab} v_R^2 M_{\rm SUSY}/M_{\rm Pl}^2
\sim 10^{-8}M_{\rm SUSY}$.  Clearly, this is very small correction to the
magnitude of $\mu$.  Suppose that quantum gravity does not respect parity
symmetry.  The coefficients $\kappa_{ab}$ will then be non--hermitian.  The
phase of the $\mu$ term will then be arg($\mu) \sim 10^{-8}$.  Through
the gluino diagram this will lead to $\bar{\theta} \sim 10^{-10}$, 
which is consistent
with $d_n$ limit.  This shows that the complex couplings $\kappa_{ab}$ can be of
order one.

The quark mass matrices can also have corrections
from Planck scale physics.  The most relevant term is $W \supset Q Q^c
\Phi \chi^c
\bar{\chi^c}/M_{\rm Pl}^2$, which will induce corrections of order
$10^{-8}v_{wk}$
for some of the quark masses.  The matrix structure need not be hermitian if
quantum gravity violates parity.  We suspect that gauged flavor symmetries
(discrete or continuous) must exist in the underlying theory, or else the
light fermion masses can become too large from quantum gravity.  Most likely the
estimate of $10^{-8} v_{wk}$ will apply for the third generation.  
If the coefficient of this non--renomralizable operator is of order $10^{-1}-
10^{-2}$, the solution to the strong CP problem via parity will be preserved.
Note that the superpotential coupling
$W \supset Q \chi Q^c \chi^c/M_{\rm Pl}$ is not invariant
under the discrete $Z_4$, unless accompanied by another factor
$S/M_{\rm Pl}$.  The correction to the quark mass matrix
from this term is extremely small $\Delta M_u \sim (M_{\rm SUSY}/M_{\rm Pl})^2
v_{wk}$.  

We have verified that all other Planck induced corrections are much below the
experimental limits on $\bar{\theta}$.

Question can be raised as to the form of SUSY breaking parameters and if they
indeed will respect parity symmetry.  A complete answer to this will have to
await a full understanding of non--perturbative SUSY breaking which is lacking
at the moment.  We note that perturbative gravity, which is utilized in
conventional supergravity models of SUSY breaking may well respect parity --
we have given an example in Polonyi model.  A second example is gauge mediated
SUSY breaking.  If SUSY is broken at a scale of $10^4-10^5$ GeV, quantum
gravity corrections for the $\mu$ term and the $A$ term,
which will be of order gravitino mass, will
of order $10^{-10}-10^{-8}$ GeV.  Even if they
are complex and non--hermitian, the strong CP problem will be solved,
as the induced $\bar{\theta}$ will be of order $10^{-10}-10^{-12}$.  We
may use one of the other proposed solutions to generate a $\mu$ term of the
weak scale in this case \cite{bm}.  
If the messenger fields do not couple to the fields $\chi^c,
\bar{\chi^c}$, they will not feel the effects of parity breaking,
although parity is broken at $v_R \sim 10^{16}$ GeV.  The effective
SUSY breaking parameters will then obey the constraints of parity.

The $d=5$ baryon number violating
operator $QQQL/M_{\rm Pl}$ in the superpotential
is forbidden in this model by $Z_4$ symmetry, but the operator $QQQLS/M_{\rm Pl}^2$
is allowed.  If the associated couplings are order one for the light generations,
we estimate proton lifetime induced by these operators to be $\tau_p \sim
10^{60}$ yr.

\section{Conclusions}

We have shown in this paper that it is possible to embed the supersymmetric
Standard Model into a parity--symmetric framework at a unification scale of
$2 \times 10^{16}$ GeV in a simple way.  Such an extension is well motivated
by the data on neutrino oscillations as well as gauge coupling unification.
We have demonstrated that this embedding can naturally solve the strong
CP problem and the SUSY phase problem simultaneously.  The effective low
energy theory is the MSSM, but with naturally small phases for the SUSY
breaking parameters along with order one phase in the CKM matrix.  
Thus it allows for large SUSY contributions to the
muon $g-2$, as indicated by experiment, without violating the bounds on
the electron EDM.  The induced $\bar{\theta}$ in these models depends strongly
on the way SUSY breaking is communicated.  With universality of squark masses
and proportionality of the $A$ terms, we found $\bar{\theta} \leq 10^{-16}$,
while with maximal deviation  from universality and proportionality consistent
with FCNC constraints $\bar{\theta} \sim
10^{-10}-10^{-8}$.  In the latter
case, the neutron EDM should be soon accessible, while $d_e$ will be much
smaller than the present experimental limit.  We have also shown that
potential corrections induced by quantum gravity are under control
in this class of models.  Since left--right gauge symmetry is realized
at a scale $v_R \sim 10^{16}$ GeV, evolution of couplings between $v_R$ and
$M_{\rm Pl}$ can induce flavor changing neutral current processes which are
in the interesting range for current and future experiments.  We plan to
study this issue in detail in a forthcoming publication.

\section*{Acknowledgements}

The work of KSB has been  supported in part by DOE Grant 
\# DE-FG03-98ER-41076, a grant from the Research Corporation,
DOE Grant \# DE-FG02-01ER4864 and by the OSU Environmental 
Institute. BD is supported by the National Science Foundation
Grant No. PHY-0070964, RNM is supported by the National
Science Foundation Grant No. PHY-9802551.

\end{document}